\documentclass[twocolumn,aps,floatfix,superscriptaddress,longbibliography]{revtex4-2}
\pdfoutput=1 %This is for the ArXiv submission only
\usepackage{amsmath,amssymb,eucal,graphicx,float,epstopdf}
\usepackage{epsfig}
\usepackage[utf8]{inputenc}

\usepackage[colorlinks=true, urlcolor=blue, anchorcolor=blue, citecolor=blue,filecolor=blue,linkcolor=blue,menucolor=blue,pagecolor=blue]{hyperref}

\begin{document}
\title{The Winner Takes It All}

\author{P.~L.~Krapivsky}
\affiliation{Department of Physics, Boston University, Boston, Massachusetts 02215, USA}
\affiliation{Santa Fe Institute, Santa Fe, New Mexico 87501, USA}

\begin{abstract} 
The winner-takes-all process takes place on an arbitrary graph. There is an agent on each vertex of the graph, and active agents at neighboring vertices play games. In each game, a randomly chosen agent wins, while the loser is eliminated from subsequent games. The games are played at random times, finishing instantaneously, and ceasing when each active agent has only losers among its neighbors. On the one-dimensional lattice, the fraction of winners in the final state is $e^{-1}$; we also determine the fractions $w_j$ of winners  who won $j=0, 1, 2$ games. For finite segments, we determine statistics of the total number of winners (the average, the variance, and all higher cumulants), the probabilities of attaining the minimum or maximum possible number of winners, and establish the behavior near the boundaries. For infinite regular trees with vertices of degree $d$, i.e., Bethe lattices with coordination number $d$, we show that the fraction of winners is $(2/d)^{d/(d-2)}$.  
\end{abstract}

\maketitle

\section{The Model}
\label{sec:WM}

We study a stochastic process involving games between agents on networks, more precisely, on connected simple graphs. (A simple graph is an undirected graph \cite{Diestel} without multiple links and loops.) A pair of agents at neighboring vertices is randomly chosen, and a randomly chosen agent wins, while the loser is eliminated from subsequent games. The process resembles 
\begin{align*}
&\textit{The winner takes it all}\\
&\textit{The loser has to fall}
\end{align*}
lyrics from the ABBA song, so we call it the winner-takes-all (WTA) process. Eventually, every active agent has only losers among the neighbors. The process stops, and remaining active agents are declared winners, although some winners haven't played a single game. The fraction of winners cannot exceed one-half on hypercubic lattices $\mathbb{Z}^d$. We will show that the WTA process is tractable on the one-dimensional lattice where the fraction of winners is $w=e^{-1}$, while the fraction of lucky winners who have not won a single game is $w_0=(2e^{-1/2}-1)^2$. 

From a game-theoretical perspective, the WTA process is elementary. Existing models of combinatorial games \cite{Conway-game} and games motivated by  economic behavior \cite{Neumann} typically involve a small number of players and often neglect spatial structure. In contrast, the WTA process emphasizes spatial interactions and can involve a large, or even infinite, number of agents. Stochastic games with many agents have been investigated mostly on complete graphs \cite{Zermelo,Bonabeau, satya:economy,ES,KK,BN1,BN2,BN3,TR08,Castellano09,TR11,BN13,TR17,Chetrite} and therefore do not incorporate spatial structure. Some of these models can be extended to arbitrary graphs \cite{satya:economy,TR17,Chetrite}. 

The exact statistics of the number of winners are analytically intractable on almost all graphs. Among a few exceptions are complete graphs where a single winner inevitably emerges. Complete bipartite graphs are also tractable. For the complete bipartite graph $K_{m,n}$, the number of active agents in one of the two parts decreases by one after each game; the evolution stops when only losers remain in one part. Star graphs form a tractable family of sparse networks. On star graphs, the central agent participates in all games, so the probability that the central agent wins $n-1$ games in a row and loses the following game is $2^{-n}$ and the number of losers in this case is $n$. The typical number of losers remains small even for large star graphs: For the star graph $K_{1,N}$, the average number of losers is $2-2^{1-N}$. 

We investigate the WTA process on one-dimensional chain graphs: The one-dimensional lattice $\mathbb{Z}$, the semi-infinite one-dimensional lattice $\mathbb{Z}_+=\{n\in \mathbb{Z}, n\geq 1\}$, and finite segments. The WTA game on such graphs is non-trivial yet still tractable. The analysis relies on methods developed in the context of random sequential adsorption (RSA); see, e.g., \cite{Flory39,Keller62,Hilhorst,Hemmer,Talbot93,Krapivsky20} and \cite{Evans93,Talbot00,KRB} for review. The analytical methods used in these studies are well-suited to one dimension, and they have been extended to other one-dimensional systems such as spin chains with zero-temperature Kawasaki dynamics, kinetically constrained Ising models, etc.; see, e.g., \cite{binary-Frisch85,binary-Frisch87,binary-Privman, binary93,binary-PK, binary-Brey,binary-JM02,binary-JM05} and \cite{JM23a} for review. 

The RSA processes are analytically intractable on higher-dimensional lattices \cite{Evans93,Talbot00,KRB}, e.g., on hypercubic lattices $\mathbb{Z}^d$ with $d\geq 2$. The same applies to the WTA process on $\mathbb{Z}^d$ and more generally on networks with numerous cycles. Analytical progress might be feasible on ladders and Bethe lattices, maybe even on the large Erd\H{o}s-R{\'e}nyi random graphs which consist predominantly of trees  and, in the percolating phase, a giant component which is locally tree-like. Indeed, the simplest RSA processes are tractable on some of these graphs \cite{RSA-ER,Pippenger,Percus91, ladder92,ladder15,ladder24,RSA-ER1,RSA-ER2,RSA-ER3}. Analyzing the RSA processes on these more complicated classes of trees and tree-like graphs is significantly more involved. Below, we show that the WTA process is tractable on two classes of infinite trees, namely, Bethe lattices and wedge sums of semi-infinite one-dimensional lattices: $\mathbb{Z}_+\vee \cdots \vee \mathbb{Z}_+$. 

In Sec.~\ref{sec:final}, we present a few basic results characterizing the final states on an infinite one-dimensional lattice. The derivations of these results are given in the following sections. In Secs.~\ref{sec:evolution}--\ref{sec:one-sided}, we describe the evolution of the WTA processes. In Sec.~\ref{sec:evolution}, we analyze the WTA process on $\mathbb{Z}$ and derive exact results for the density of clusters of agents who haven't yet lost a game. In Sec.~\ref{sec:corr}, we probe correlations. Specifically, we compute the connected equal-time pair correlation function. The evolution of the WTA process on $\mathbb{Z}_+$ is the subject of Sec.~\ref{sec:one-sided}. In particular, we compute the probability that the agent on the boundary is a lucky winner and the probability of winning a single game. Using these results, we derive  the fractions $w_j$ of winners on $\mathbb{Z}$ who won $j=0, 1, 2$ games.

In Secs.~\ref{sec:segment}--\ref{sec:FCS}, we consider the final states of the WTA process on segments. (The same methods apply to rings; the asymptotic behaviors are the same as for segments.) In Sec.~\ref{sec:segment}, we compute the average number of winners on the segment. We also determine the probability that the agent on the boundary is the winner and, more generally, the probabilities that this agent has not played a game or has participated in exactly one game. In Sec.~\ref{sec:FCS}, we compute the full counting statistics for the number of winners, namely, we determine the cumulant generating function. The computation proceeds along the same lines as that for the average number of winners. We also determine the probabilities of reaching the final state with the minimal or maximal number of winners. In Sec.~\ref{sec:Bethe}, we briefly consider the WTA process on Bethe lattices. A few challenges for future work are presented in Sec.~\ref{sec:disc}.

\section{One-dimensional Lattice: Final States }
\label{sec:final}

The WTA process can be played on any simple graph. Since different connected components of a graph do not interact during the WTA process, we can assume that the graph is connected. In what follows, we restrict attention to connected trees. In this section, we summarize several results concerning the final states of the WTA process on the one-dimensional lattice. The derivations of these and other results, particularly those concerning the evolution, are presented in the following sections. 

A final configuration looks like
\begin{equation}
\label{jammed}
\ldots \circ\,\bullet\,\circ\,\circ\,\bullet\,\circ\,\circ\,\bullet\,\circ\,\bullet\,\circ\,\circ\,\circ\,\circ\,\bullet\,\circ\,\circ\,\bullet\,\circ\ldots
\end{equation}
Here $\circ$ denotes the loser, and $\bullet$ denotes the winner. Winners are separated by a string of losers. Let 
\begin{equation}
\label{Pi-s:def}
\Pi_s = \text{Prob}\big[\bullet \underbrace{\circ \cdots \circ}_{s} \bullet\big]
\end{equation}
be the frequency of segments with $s$ losers between adjacent winners. The sum rule
\begin{equation}
1 = \sum_{s\geq 1}(s+1)\Pi_s
\end{equation}
expresses the decomposition of the lattice into isolated winners and intervening loser domains. The fraction of winners 
\begin{equation}
\label{winners}
\sum_{s\geq 1}\Pi_s = e^{-1}=0.36787944117\ldots
\end{equation}
is computed in Sec.~\ref{sec:evolution}. The derivation directly probes the fraction of winners. Computing the domain size distribution $\Pi_s$ requires more elaborate treatment presented in Sec.~\ref{subsec:domain} where we derive
\begin{equation}
\label{Pi_s}
\Pi_s = e^{-1}\left[\frac{1}{s!}-\frac{1}{(s+1)!}\right]
\end{equation}

Each winner participated in a certain number of games and won them all. Generally, an agent with $d$ neighbors plays at most $d$ games. A winner has participated in $j \leq 2$ games on $\mathbb{Z}$. Let $w_0, w_1, w_2$ be the fractions of such agents in the final state. We have
\begin{subequations}
\label{sum-rules}
\begin{equation}
\label{winners:frac}
w_0 + w_1 + w_2 = e^{-1}
\end{equation}
One can similarly divide the losers into those who lost in the first or second game they played. If $\ell_1$ and $\ell_2$ are the fractions of such agents,
\begin{equation}
\label{losers:frac}
\ell_1 + \ell_2 = 1 - e^{-1}
\end{equation}
On average, an agent participates in $w_1 + 2w_2 + \ell_1 + 2\ell_2$ games. Taking into account that the average number of games per bond is $1 - e^{-1}$ and two players participate in each game, we conclude that 
\begin{equation}
\label{sum}
w_1 + 2w_2 + \ell_1 + 2\ell_2 = 2(1 - e^{-1})
\end{equation}
\end{subequations}

Direct calculation (Sec.~\ref{subsec:boundary}) yields the fractions of winners 
\begin{subequations}
\label{winners:012}
\begin{align}
\label{w0}
w_0 &= 1-4e^{-\frac{1}{2}}+4e^{-1}\\
\label{w1}
w_1 &= -2+6e^{-\frac{1}{2}}-4e^{-1}\\
\label{w2}
w_2 &= 1-2e^{-\frac{1}{2}}+e^{-1}
\end{align}
\end{subequations}
Combining \eqref{winners:012} and the sum rules \eqref{losers:frac}--\eqref{sum} we obtain the fractions of losers
\begin{subequations}
\begin{align}
\label{L1}
\ell_1 &= 2e^{-\frac{1}{2}}-2e^{-1}\\
\label{L2}
\ell_2 &= 1-2e^{-\frac{1}{2}}+e^{-1}
\end{align}
\end{subequations}
Interestingly, $w_2=\ell_2$. It would be useful to find a direct derivation of this identity. 

We also computed (Sec.~\ref{subsec:pair}) the connected pair correlation function 
\begin{equation}
\label{Cj:correl}
C_j = \langle\!\langle O_0 O_j\rangle\! \rangle = \langle O_0 O_j\rangle -  \langle O_0\rangle  \langle O_j\rangle
\end{equation}
Here $O_j$ is the indicator function for the occupancy by active agents, i.e., $O_j=1$ if the agent at site $j$ is active and $O_j=0$ otherwise. In the final state
\begin{equation}
\label{Cj:final}
C_j = - e^{-1}\sum_{i\geq j+1}\frac{(-1)^i}{i!}
\end{equation}
from which
\begin{equation}
\label{Cj:asymp}
C_j \simeq e^{-1}\frac{(-1)^j}{(j+1)!}\quad\text{as}\quad j\to \infty
\end{equation}

The sum of $C_j$ over $j\in \mathbb{Z}$ satisfies 
\begin{equation}
\label{Cj:sum-var}
\sum_{j=-\infty}^\infty C_j = \lim_{L\to\infty}L^{-1}  \langle\!\langle N^2\rangle\! \rangle
\end{equation}
following from the definitions of the connected pair correlation function and the variance $\langle\!\langle N^2\rangle\! \rangle$ of the 
number of losers (equivalently, active agents) on a long segment. The relation between the connected pair correlation function and the variance is very general, and \eqref{Cj:sum-var} has been used \cite{KRB} in the context of RSA. 

We established the asymptotic behavior of the variance of the number of losers in the final state on long segments [see Eq.~\eqref{var:lim}], from which 
\begin{equation}
\label{Cj:sum}
\sum_{j=-\infty}^\infty C_j = 3 e^{-2} -  e^{-1}
\end{equation}
Using \eqref{Cj:final} we find
\begin{equation}
\label{Cj:sum-plus}
\sum_{j\geq 1} C_j = 2 e^{-2} -  e^{-1}
\end{equation}
Using $C_0=e^{-1}-e^{-2}$ following from \eqref{Cj:final} and the symmetry $C_j=C_{-j}$ we find that \eqref{Cj:sum} and \eqref{Cj:sum-plus} are equivalent. 

The pair correlation function \eqref{Cj:final} was established using a dynamical treatment of the WTA process on the one-dimensional lattice, while the variance in \eqref{Cj:sum-var} was computed via a static analysis of the WTA process on long segments. The agreement of these different computations provides a useful consistency check.  

\section{Dynamics of the WTA process on $\mathbb{Z}$}
\label{sec:evolution}

Even if we only want to describe the average characteristics of the final (jammed) state, experience with RSA processes suggests  treating the process dynamically; see, e.g., \cite{Keller62} and  Refs.~\cite{Evans93,Talbot00,KRB} for a review. 

Each bond is chosen independently with the same rate, which we set to unity. The game between the agents adjacent to the bond is not played if at least one of the two agents is already the loser; otherwise, one randomly chosen agent becomes the loser. Let 
\begin{equation}
\label{winners:seg}
\mathcal{W}_n(t) = \text{Prob}\big[\circ \underbrace{\bullet \cdots \bullet}_{n} \circ\big]
\end{equation}
be the density of segments with exactly $n$ active agents at time $t$. These quantities evolve according to 
\begin{equation}
\label{ME:winners}
\frac{d\mathcal{W}_n}{dt} =-(n-1)\mathcal{W}_n + \mathcal{W}_{n+1}+2\sum_{j\geq n+2}\mathcal{W}_j
\end{equation}
Indeed, each of the $n-1$ bonds between the winners in segment \eqref{winners:seg} could be chosen, explaining the loss term in \eqref{ME:winners}. The gain term $\mathcal{W}_{n+1}$ accounts for boundary agents in the segment of length $n+1$ losing in games; the following gain terms arise when an agent in the bulk loses a game. 

To solve an infinite set of master equations \eqref{ME:winners}, we employ an exponential ansatz:
\begin{equation}
\label{exp:winners}
\mathcal{W}_n = B b^{n-1}
\end{equation}
Plugging \eqref{exp:winners} into \eqref{ME:winners}, we reduce an infinite system of ordinary differential equations \eqref{ME:winners} into two coupled differential equations
\begin{subequations}
\begin{align}
\label{a:eq}
&\frac{db}{dt} = - b\\
\label{A:eq}
&B^{-1}\,\frac{dB}{dt} =  b+\frac{2b^2}{1-b}
\end{align}
\end{subequations}
Treating $B$ as a function of $b$ rather than time we recast \eqref{A:eq} into
\begin{equation}
\label{Aa:eq}
B^{-1}\,\frac{dB}{db}  =  1-\frac{2}{1-b}
\end{equation}
from which $B = C(1-b)^2 e^{b-1}$. The initial condition 
\begin{subequations}
\label{IC-W}
\begin{align}
\label{IC-n}
& \mathcal{W}_n(0)=0   \qquad (n\geq 1)\\
\label{IC-sum}
& \sum_{n\geq 1} n \mathcal{W}_n(0) = 1
\end{align}
\end{subequations}
with the latter one ensuring that all agents are initially active. The initial condition \eqref{IC-n} implies $B(t=0)=0$ which together with $B = C(1-b)^2 e^{b-1}$
fixes $b(0)=1$ and the amplitude $C=1$. Integrating \eqref{a:eq} subject to $b(0)=1$ we obtain $b = e^{-t}$. Gathering all these results we arrive at 
\begin{equation}
\label{Wn:sol}
\mathcal{W}_n(t) = \tau^2 e^{-\tau} (1-\tau)^{n-1}
\end{equation}
In \eqref{Wn:sol} and many following equations, we often use a modified time variable
\begin{equation}
\tau = 1-e^{-t}
\end{equation}
Thus, $\mathcal{W}_1(\infty)= e^{-1}$ and $\mathcal{W}_n(\infty)=0$ for $n>1$, justifying \eqref{jammed} and confirming that the fraction of winners is indeed given by \eqref{winners}. The fraction of agents who have not lost a game up to time $t$ is
\begin{equation}
\label{winners-t}
w(t)=\sum_{n\geq 1}n\mathcal{W}_n(t)= e^{-\tau}
\end{equation}

Another natural quantity is the density 
\begin{equation}
\label{winners:seg+}
W_n(t) = \text{Prob}\big[\underbrace{\bullet \cdots \bullet}_{n}\big]
\end{equation}
of segments with $n$ consecutive active agents at time $t$; in contrast to \eqref{winners:seg}, the states of the two borderline agents are unspecified. The densities $W_n$ evolve according to the master equation
\begin{equation}
\label{ME:simple}
\frac{dW_n}{dt} =-(n-1)W_n - W_{n+1},
\end{equation}
which is simpler than the master equation \eqref{ME:winners} as there are no gain terms. The initial condition $W_n(0)=1$ for all $n\geq 1$ is also simpler than the rather subtle initial condition \eqref{IC-W}. An exponential ansatz is again applicable, and the solution to \eqref{ME:simple} reads  
\begin{equation}
\label{Wn}
W_n(t) = e^{-\tau} (1-\tau)^{n-1}
\end{equation}
The general relation 
\begin{equation}
\label{Wn:W}
\mathcal{W}_n = W_n - 2 W_{n+1}+W_{n+2}
\end{equation}
agrees with \eqref{Wn:sol} and \eqref{Wn}.

\section{Correlations}
\label{sec:corr}

The degree to which the spatial arrangement of active agents deviates from randomness can be quantified by various correlation functions. One natural characteristic is the connected equal-time pair correlation function 
\begin{equation}
\label{Cj:def}
C_j = \langle\!\langle O_0O_j\rangle\!\rangle= \langle O_0 O_j\rangle -  \langle O_0\rangle  \langle O_j\rangle
\end{equation}
where $O_j$ is the indicator function for the occupancy of site $j$:
\begin{equation}
O_j =
\begin{cases}
1 &\text{if the agent at site $j$ is active}\\
0 & \text{otherwise}
\end{cases}
\end{equation}

Another natural characteristic of the emerging spatial arrangement is provided by the distribution of domains of losers \eqref{Pi-s:def}. In the final state, the quantities $C_j$ and $\Pi_s$ do not satisfy close relations. To overcome this obstacle, we introduce significantly more involved infinite sets of auxiliary quantities that satisfy closed sets of evolutionary equations. Using exponential ans\"{a}tze resembling \eqref{exp:winners}, we solve these equations and extract $C_j$ and $\Pi_s$ in the $t\to\infty$ limit. This trick goes back to Ref.~\cite{Talbot93} where it was applied to solving the ballistic-deposition problem. 

\subsection{Pair correlation function}
\label{subsec:pair}

On the one-dimensional lattice, the average occupancy is spatially homogeneous, $\langle O\rangle =e^{-\tau}$, and Eq.~\eqref{Cj:def} becomes
\begin{equation}
\label{Cj-Pj}
C_j = P_j - e^{-2\tau}, \qquad P_j = \langle O_0 O_j\rangle
\end{equation}
To avoid cluttering formulas, we suppress the time variable when it does not lead to confusion.

The computations of $C_j$ have been successfully performed for a few simplest one-dimensional RSA processes \cite{Hilhorst,Hemmer}; see also \cite{Evans93, Talbot00,KRB}. For the WTA process, we already know 
\begin{subequations}
\begin{align}
\label{C0}
C_0 & = \langle O\rangle - \langle O\rangle^2 \\
\label{C1}
C_1 & = e^{-t} \langle O\rangle - \langle O\rangle^2
\end{align}
\end{subequations}
To derive \eqref{C0}, we use the definition $C_0=\langle O^2\rangle-\langle O\rangle^2$ and identity $O^2\equiv O$ valid for any binary $\{0,1\}$ variable. Using $\langle O_0 O_1\rangle =W_2=e^{-t}\langle O\rangle$ we establish \eqref{C1}. 

In the final state, Eqs.~\eqref{C0}--\eqref{C1} become
\begin{align}
\label{C:01}
C_0 = e^{-1} - e^{-2}, \qquad  C_1 = - e^{-2}
\end{align}

To derive $C_j$ for arbitrary $j$, we introduce the probabilities
\begin{equation}
\label{win-x-win}
W_{m,j,n} = \text{Prob}\big[\underbrace{\bullet \cdots \bullet}_{m}\,\underbrace{\star \cdots \star}_{j-1}\,\underbrace{\bullet \cdots \bullet}_{n}\big]
\end{equation}
of configurations with $m\geq 1$ active agents on the left and $n\geq 1$ active agents on the right surrounding $j-1$ sites with agents in unspecified state (these agents are denoted by $\star$). We want to compute
\begin{equation}
\label{Pj:def}
P_j = W_{1,j,1} = \text{Prob}\big[\bullet\,\underbrace{\star \cdots \star}_{j-1}\,\bullet \big]
\end{equation}
but we need a much more general class of configurations \eqref{win-x-win}. The probabilities $W_{m,j,n}$ satisfy 
\begin{eqnarray}
\label{ME:mjn}
&&\frac{dW_{m,j,n}}{dt} = -(m-1)W_{m,j,n} - \frac{W_{m+1,j,n}+W_{m+1,j-1,n}}{2} \nonumber \\
& & - (n-1)W_{m,j,n} - \frac{W_{m,j,n+1}+W_{m,j-1,n+1}}{2}
\end{eqnarray}
derived similarly to \eqref{ME:winners}. The exponential ansatz 
\begin{equation}
W_{m,j,n}(t) = e^{-(m+n-2)t}P_j(t)
\end{equation}
is consistent with \eqref{Pj:def}, and it reduces the master equations \eqref{ME:mjn} to
\begin{equation}
\label{Pj:eq}
\frac{dP_j}{dt} = -e^{-t}\big[P_j+P_{j-1}\big]
\end{equation}
Using the modified time variable and the auxiliary variable $\widehat{P}_j=P_j e^\tau$ we simplify \eqref{Pj:eq} to 
\begin{equation}
\label{Qj:eq}
\frac{d\widehat{P}_j}{d\tau} = -\widehat{P}_{j-1}
\end{equation}
Since $P_1=W_2=(1-\tau)e^{-\tau}$, we have $\widehat{P}_1=1-\tau$, and hence Eqs.~\eqref{Qj:eq} are applicable for all $j\geq 1$ if we set $\widehat{P}_0=1$. The solution to the recurrence \eqref{Qj:eq} subject to the initial condition $\widehat{P}_j(0)=1$ for all $j\geq 1$ can be found using, e.g., the generating function technique. One gets
\begin{equation}
\label{Qj:sol}
\widehat{P}_j = \sum_{i=0}^j\frac{(-\tau)^i}{i!} = e^{-\tau}- \sum_{i\geq j+1}\frac{(-\tau)^i}{i!}
\end{equation}
for $j\geq 0$. One can also prove \eqref{Qj:sol} by induction. Thus
\begin{equation}
\label{Pj:sol}
P_j =  e^{-2\tau} - e^{-\tau}\sum_{i\geq j+1}\frac{(-\tau)^i}{i!}
\end{equation}
Combining \eqref{Cj-Pj} and \eqref{Pj:sol} we obtain
\begin{equation}
\label{Cj:sol}
C_j = - e^{-\tau}\sum_{i\geq j+1}\frac{(-\tau)^i}{i!}
\end{equation}
Specializing \eqref{Cj:sol} to the final state, $\tau=1$, we arrive at the announced result \eqref{Cj:final}.

\subsection{Domain size distribution}
\label{subsec:domain}

In the final state, domains of winners have length one as winners cannot be neighbors, while the size distribution of domains of losers is nontrivial. To determine $\Pi_s$, we consider the evolution of 
\begin{equation}
\label{w-losers-w}
P_s(m,n; t) = \text{Prob}\big[\underbrace{\bullet \cdots \bullet}_{m} \underbrace{\circ \cdots \circ}_{s}\underbrace{\bullet \cdots \bullet}_{n}\big]
\end{equation}
In the final state, the winners cannot be neighbors, so the only nonvanishing densities are $\Pi_s=P_s(1,1; \infty)$. 

The probabilities $P_s(m,n; t)$ evolve according to
\begin{eqnarray}
\label{P-mns}
\frac{dP_s(m,n)}{dt} &=& - (m+n-2) P_s(m,n) \nonumber \\
&-&\frac{P_s(m+1,n)+P_s(m,n+1)}{2}\nonumber \\
&+&[1-\delta_{s,1}]\frac{P_{s-1}(m+1,n)+ P_{s-1}(m,n+1)}{2}\nonumber \\
&+& \delta_{s,1}W_{m+1+n}
\end{eqnarray}
The specificity of the domain of losers of the minimal width is reflected by Kronecker delta symbol $\delta_{s,1}$ in the gain terms. The form of master equations \eqref{P-mns} suggest to seek the solution in the form
\begin{equation}
\label{Rs:def}
P_s(m,n; t) = e^{-(m+n-2)t} R_s(t)
\end{equation}
This ansatz reduces Eqs.~\eqref{P-mns}  to  
\begin{subequations}
\label{R-eq}
\begin{align}
\label{R1}
\frac{d R_1}{d\tau} & =  - R_1+(1-\tau)e^{-\tau}      &  s=1 \\
\label{Rs}
\frac{d R_s}{d \tau} &=  - R_s + R_{s-1}                 & s\geq 2
\end{align}
\end{subequations}
Introducing the auxiliary variables $\widehat{R}_s=R_s e^\tau$ we reduce Eqs.~\eqref{R-eq} to 
\begin{equation}
\label{R:eq}
\frac{d\widehat{R}_s}{d\tau} = \widehat{R}_{s-1}
\end{equation}
valid for all $s\geq 1$ if we set $\widehat{R}_0=1-\tau$. Solving \eqref{R:eq} subject to the initial condition $\widehat{R}_s(0)=1$ for all $s\geq 1$ we obtain
\begin{equation}
\label{Rs:sol}
\widehat{R}_s = \frac{\tau^s}{s!}-\frac{\tau^{s+1}}{(s+1)!}
\end{equation}
Collecting previous results we arrive at
\begin{equation}
\label{Pmns:sol}
P_s(m,n; t) = e^{-(m+n-2)t-\tau}\left[\frac{\tau^s}{s!}-\frac{\tau^{s+1}}{(s+1)!}\right]
\end{equation}
The time-dependent distribution of domains of losers 
\begin{equation}
\label{Ps:sol}
P_s(1,1; t) = e^{-\tau}\left[\frac{\tau^s}{s!}-\frac{\tau^{s+1}}{(s+1)!}\right]
\end{equation}
reduces to the announced domain size distribution \eqref{Pi_s} in the final state.

\section{Dynamics of the WTA process on $\mathbb{Z}_+$}
\label{sec:one-sided}

Here we consider the WTA process on the semi-infinite lattice $\mathbb{Z}_+=\{j\in \mathbb{Z}, j\geq 1\}$. The full description is significantly more challenging than the description of the WTA process on $\mathbb{Z}$ due to spatial inhomogeneity. On the infinite one-dimensional lattice, it was technically easier to deal with probabilities $W_n$ than with probabilities $\mathcal{W}_n$ with specified fates of agents beyond the cluster of $n$ agents. The same is true for the semi-infinite lattice. Because we are mostly interested in agents near the boundary, the simplest analogs of $W_n$ are the probabilities
\begin{equation}
\label{winners:B}
B_n(t) = \text{Prob}\big[\underbrace{\blacktriangleleft\bullet \cdots \bullet}_{n}\big]
\end{equation}
specifying that $n$ consecutive agents at the boundary of $\mathbb{Z}_+$ are active, but leaving unspecified the state of the $(n+1)^{\text{st}}$ agent. (We denote by $\blacktriangleleft$ an active agent on the boundary.) The probabilities $B_n$ satisfy 
\begin{equation}
\label{B:winners}
\frac{dB_n}{dt} =-(n-1)B_n - \frac{1}{2}\,B_{n+1}
\end{equation}
derived similarly to Eqs.~\eqref{ME:simple}. We use again an exponential ansatz, $B_n = B b^{n-1}$, and reduce an infinite system \eqref{B:winners} to a pair of differential equations
\begin{align*}
\frac{db}{dt} = - b, \qquad B^{-1}\,\frac{dB}{dt} =  \frac{b}{2}
\end{align*}
which we solve as in Sec.~\ref{sec:evolution} and find $b = e^{-t}=1-\tau$ as before, while the amplitude is now $B=e^{-\tau/2}$. Thus
\begin{equation}
\label{Bn:sol}
B_n(t) = e^{-\tau/2} (1-\tau)^{n-1}
\end{equation}
The probability that the first agent is active at time $t$ is
\begin{equation}
\label{w1:sol}
w^{(1)}(t) =B_1(t) = e^{-\tau/2} 
\end{equation}
The probability that the first agent is the winner is
\begin{equation}
\label{w1:final}
w^{(1)} = \frac{1}{\sqrt{e}}
\end{equation}
since $t=\infty$ corresponds to $\tau=1$. 

The fate of the second agent is captured by $B_2$ which we already know and the probability $C_1$ of the pattern $\vartriangleleft \bullet$, where $\vartriangleleft$ denotes an agent on the boundary that lost the game during the time interval $(0,t)$. To determine $C_1$, we need again an infinite set of probabilities 
\begin{equation}
\label{winners:C}
C_n(t) = \text{Prob}\big[\vartriangleleft \underbrace{\bullet \cdots \bullet}_{n}\big]
\end{equation}
since the evolution of $C_1$ is coupled with $C_2$, which in turn is coupled with $C_3$, etc.  In \eqref{winners:C}, the state of the $(n+2)^{\text{nd}}$ agent is unspecified. The probabilities $C_n$ satisfy 
\begin{equation}
\label{C:winners}
\frac{dC_n}{dt} =\frac{1}{2}\,B_{n+1}-(n-1)C_n - \frac{1}{2}\,C_{n+1}
\end{equation}
The homogeneous part of the linear inhomogeneous differential equation \eqref{C:winners} coincides with the master equation \eqref{B:winners} for $B_n$. We thus seek a solution by the variation of parameters: $C_n = FB_n$. We recover the same results $b = e^{-t}$ and $B=e^{(b-1)/2}$ as before, together with $\frac{dF}{db}=-\frac{1}{2}$, from which $F=\frac{1-b}{2}$. Thus
\begin{equation}
\label{Cn:sol}
C_n(t) = \frac{1}{2}\,\tau\,e^{-\tau/2} (1-\tau)^{n-1}
\end{equation}
The second agent (the first is the agent at the boundary) is the winner with probability $w^{(2)} = C_1(\infty)$. Combining this relation with \eqref{Cn:sol}, we arrive at  
\begin{equation}
\label{w2:final}
w^{(2)} = \frac{1}{2\sqrt{e}}
\end{equation}
More generally, the second agent is active at time $t$ with probability
\begin{equation}
\label{w2:sol}
w^{(2)}(t) = B_2(t)+C_1(t) =  \left(1-\frac{\tau}{2}\right)e^{-\tau/2} 
\end{equation}

The fate of the third agent is captured by $B_3$ and $C_2$, which we already know, and by the probabilities of the patterns $\circ \circ \bullet$ and $\bullet \circ \bullet$. To determine the latter probabilities, we introduce two infinite sets of probabilities, $D_n(t)$ and $\overline{D}_n(t)$, corresponding to the patterns
\begin{subequations}
\begin{align}
D_n(t) &= \text{Prob}\big[\vartriangleleft \circ \underbrace{\bullet \cdots \bullet}_{n}\big] \\
\overline{D}_n(t) & = \text{Prob}\big[\blacktriangleleft \circ \underbrace{\bullet \cdots \bullet}_{n} \big]
\end{align}
\end{subequations}
The master equations describing the evolution of these probabilities read
\begin{subequations}
\begin{align}
\frac{dD_n}{dt} &= \frac{1}{2}\,C_{n+1}-(n-1)D_n - \frac{1}{2}\,D_{n+1}\\
\frac{d \overline{D}_n}{dt} &= B_{n+2}-(n-1)\overline{D}_n - \frac{1}{2}\,\overline{D}_{n+1}
\end{align}
\end{subequations}
We solve these master equations by employing an exponential ansatz. The solutions take the simple form if we express them in terms of the modified time variable $\tau$:
\begin{subequations}
\label{Dn:sol}
\begin{align}
\label{Dn-1:sol}
D_n &= \frac{1}{8}\,\tau^2\, e^{-\tau/2} (1-\tau)^{n-1} \\
\label{Dn-2:sol}
\overline{D}_n & =\left(\tau-\frac{\tau^2}{2}\right)e^{-\tau/2} (1-\tau)^{n-1} 
\end{align}
\end{subequations}
The third agent is active at time $t$ with probability 
\begin{equation}
\label{w3-long}
w^{(3)}(t) =  B_3(t)+ C_2(t)+D_1(t) + \overline{D}_1(t)
\end{equation}
Using \eqref{Bn:sol}, \eqref{Cn:sol}, \eqref{Dn-1:sol}--\eqref{Dn-2:sol}, we rewrite \eqref{w3-long} as 
\begin{equation}
\label{w3:sol}
w^{(3)}(t) = \left(1-\frac{\tau}{2}+\frac{\tau^2}{8}\right)e^{-\tau/2} 
\end{equation}
from which we get 
\begin{equation}
\label{w3:final}
w^{(3)} = \frac{3}{4\sqrt{e}}
\end{equation}

Extending this direct approach to computing the probabilities $w^{(j)}$ quickly becomes cumbersome as $j$ increases. We anticipate
\begin{equation}
\label{wj:sol}
w^{(j)}(t) =  \Phi_j(\tau)\,e^{-\tau/2} 
\end{equation}
Equations \eqref{w1:sol}, \eqref{w2:sol}, and \eqref{w3:sol} give
\begin{equation}
\label{P123}
\Phi_1  = 1, \quad \Phi_2  = 1 - \frac{\tau}{2}\,, \quad 
\Phi_3  = 1 - \frac{\tau}{2} + \frac{\tau^2}{8}
\end{equation}
This suggests the following conjecture
\begin{equation}
\label{guess}
\Phi_j = \sum_{i=0}^{j-1} \frac{1}{i!}\left(-\frac{\tau}{2}\right)^i
\end{equation}
agreeing with \eqref{P123} for $j\leq 3$ and giving $\Phi_\infty=e^{-\tau/2}$, so that $w^{(\infty)}= e^{-\tau}$ which is the exact result for the infinite one-dimensional lattice. Equation \eqref{guess} implies that the $j^\text{th}$ agent is the winner with probability
\begin{equation}
\label{wj:final}
w^{(j)} =  A_j\,e^{-\frac{1}{2}}, \qquad A_j = \sum_{i=0}^{j-1} \frac{(-1)^i}{i!\,2^i}
\end{equation}
The approach to the bulk value is very fast:
\begin{equation}
\label{wj:final-large}
w^{(j)} - e^{-1} \simeq e^{-\frac{1}{2}}\,\frac{(-1)^{j-1}}{j!\,2^j}
\end{equation}
when $j\gg 1$. 

One can define a one-parameter family of infinite linear graphs parametrized by the number of arms: the graph with one arm is the semi-infinite lattice $\mathbb{Z}_+$; the graph with two arms is the infinite one-dimensional lattice $\mathbb{Z}$; etc. In Appendix \ref{ap:arms}, we show how to extend some  computations of this section to infinite linear graphs with an arbitrary number of arms.

\section{Segments}
\label{sec:segment}

Here, we consider the WTA process on segments. The dynamical approach of Sect.~\ref{sec:evolution} remains applicable, but it is preferable to employ a static approach if we are only interested in the statistics of final states. Instead of the total number of winners, we consider the dual quantity, namely, the total number of losers. 

The number  $N$ of losers on the segment with $L$ sites is a random quantity. The maximal and minimal possible numbers of losers are 
\begin{equation}
\label{min-max}
N_\text{max} = L-1, \qquad N_\text{min} = \left\lfloor \frac{L}{2}\right\rfloor
\end{equation}

Jammed configurations with maximal number of losers have a single winner located at an arbitrary position:
\begin{equation}
\label{dense}
\underbrace{\circ  \cdots \circ  \bullet \circ  \cdots \circ }_{L}
\end{equation}
There are $L$ such jammed configurations. 

When $L$ is odd, there is a single jammed configuration with minimal number of losers:
\begin{subequations}
\label{sparse}
\begin{equation}
\label{sparse:odd}
\underbrace{\bullet \circ \bullet \circ \cdots \circ \bullet\, \circ \bullet}_{L}
\end{equation}
When $L$ is even, there are two jammed configurations with minimal number of losers:
\begin{equation}
\label{sparse:even}
\underbrace{\circ \bullet \circ \bullet \cdots \circ \bullet}_{L}   \quad\text{and}\quad  
\underbrace{\bullet \circ \bullet \cdots \circ \bullet\, \circ}_{L}
\end{equation}
\end{subequations}

In Sec.~\ref{subsec:av}, we compute the average number of losers $A_L=\langle N\rangle$.  (We already know the asymptotic behavior: $L^{-1}A_L\to 1-e^{-1}$ as $L\to\infty$.)  In Sec.~\ref{subsec:boundary}, we determine the fate of the agent at the boundary, from which we deduce the announced results \eqref{winners:012}. In Sec.~\ref{subsec:min-max}, we establish the probabilities to reach jammed states with minimal and maximal numbers of losers.

\subsection{The average number of losers}
\label{subsec:av}

One can compute $A_L=\langle N\rangle$ by hand when $L$ is small. One gets $N=0$ when $L=1$ and $N=1$ when $L=2$. For $L\geq 3$, the total number $N$ of losers varies from realization to realization. For $L=3$, one finds 
\begin{equation}
\label{L=3}
N = \begin{cases}
1 & \text{prob} ~~ \frac{1}{2}\\
2 & \text{prob} ~~ \frac{1}{2}
\end{cases}
\end{equation}
Continuing one finds
\begin{equation}
\label{L=4}
N = \begin{cases}
2 & \text{prob} ~~ \frac{5}{6}\\
3 & \text{prob} ~~ \frac{1}{6}
\end{cases}
\end{equation}
for $L=4$, 
\begin{equation}
\label{L=5}
N = \begin{cases}
2 & \text{prob} ~~ \frac{1}{4}\\
3 & \text{prob} ~~ \frac{17}{24}\\
4 & \text{prob} ~~ \frac{1}{24}
\end{cases}
\end{equation}
for $L=5$, etc. leading to
\begin{equation*}
A_1=0,\quad A_2=1, \quad A_3 = \frac{3}{2}, \quad A_4 = \frac{13}{6}, \quad A_5 = \frac{67}{24}
\end{equation*}
etc. This direct enumeration becomes increasingly cumbersome as $L$ increases.
 
 To determine the general expression for $A_L$, we note that when $L\geq 2$, the first loser can be at one of the two boundary sites with probability $1/(L-1)$, and at any bulk site with probability $1/(L-1)$. We label the sites of the interval from 1 to $L$ and denote by $k$ the label of the first loser. The average number of losers $A_L$ can be determined from a recurrence
\begin{equation}
\label{AL:rec}
A_L = 1+ \frac{1}{L-1}\,A_{L-1}+\frac{1}{L-1}\sum_{k=2}^{L-1} \big(A_{k-1}+A_{L-k}\big)
\end{equation}
The first term on the right-hand side of \eqref{AL:rec} counts the first loser, and the next terms account for the average number of subsequent losers. The second term accounts for the first loser at the boundary, $k=1$ or $k=L$. The remaining terms describe the realizations when the first loser is in the bulk, $k=2,\ldots,L-1$. Using the generating function 
\begin{equation}
\mathcal{A}(x) = \sum_{L\geq 2} A_L\, x^{L-1}
\end{equation}
we convert the recurrence \eqref{AL:rec} into a differential equation
\begin{equation}
\frac{d\mathcal{A}(x)}{dx} = \frac{1+x}{1-x}\,\mathcal{A}(x)+\frac{1}{(1-x)^2}
\end{equation}
Solving this linear inhomogeneous differential equation subject to the initial condition $\mathcal{A}(0)=0$ we obtain
\begin{equation}
\label{Ax}
\mathcal{A}(x) = \frac{1-e^{-x}}{(1-x)^2}
\end{equation}
from which
\begin{equation}
\label{AL:sol}
A_L = \left(1-e^{-1}\right)L - e^{-1} - \sum_{n\geq L+1} \frac{(-1)^n}{n!}\,(n-L)
\end{equation}
The first two terms on the right-hand side of \eqref{AL:sol} provide an excellent approximation when $L\gg 1$ since the next term is $1/(L+1)!$. The average fraction of the winners is therefore 
\begin{equation}
w_L = 1-\frac{A_L}{L} = e^{-1} +  e^{-1}\,L^{-1} +\frac{(-1)^{L+1}}{L(L+1)!}+\ldots
\end{equation}
leading to the announced result \eqref{winners} and providing finite-size corrections.

The above method of computing the average goes back to Flory \cite{Flory39}, who used it in the context of irreversible reactions on a polymer chain. Similar techniques can be used for the computation of the behavior near the boundaries as we now demonstrate.

\subsection{Behavior at the boundary}
\label{subsec:boundary}

Denote by $b_L$ the probability that the agent at the boundary, say the leftmost site in the segment, is the winner. For small $L$, one can compute $b_L$ by hand:
\begin{equation}
b_1=1,\quad b_2=\frac{1}{2}\,,\quad b_3=\frac{5}{8}\,, \quad b_4 = \frac{29}{48}
\end{equation}
etc. Generally for $L\geq 2$, the probability $b_L$ satisfies the recurrence
\begin{equation}
\label{wL:rec}
b_L =  \frac{1}{2(L-1)}\sum_{k=1}^{L-1} b_k + \frac{1}{2(L-1)}\sum_{k=1}^{L-2} b_k
\end{equation}
Here $k+1$ is the first loser, with $k=1,\ldots,L-1$ if the $(k+1)$st agent lost to the neighbor on the left, and $k=1,\ldots,L-2$ if the $(k+1)$st agent lost to the neighbor on the right. 

Massaging the recurrence \eqref{wL:rec} we recast it into
\begin{equation}
\label{wL:rec-2}
Lb_L + \frac{1}{2}\,b_{L-1}  = \sum_{k=1}^{L} b_k 
\end{equation}
which is valid for $L\geq 1$ if we set $b_0=0$. 
Using the generating function 
\begin{equation}
\label{Wx:def}
\mathcal{B}(x) = \sum_{L\geq 1} b_{L}\, x^L
\end{equation}
we convert the recurrence \eqref{wL:rec-2} into a differential equation
\begin{equation}
\frac{d \mathcal{B}(x)}{dx} = \left(\frac{1}{x}+\frac{1}{1-x}-\frac{1}{2}\right) \mathcal{B}(x)
\end{equation}
which is solved to yield the generating function 
\begin{equation}
\label{Wx:sol}
\mathcal{B}(x) = \frac{x\,e^{-x/2}}{1-x}
\end{equation}
encapsulating all $b_L$. The amplitude in \eqref{Wx:sol} was fixed by the boundary condition $b_1=1$. 

Using \eqref{Wx:def} and \eqref{Wx:sol} and taking the $L\to\infty$ limit gives $b_\infty=e^{-1/2}$. Equivalently, this is the probability $w^{(1)}$ that the first agent in the semi-infinite lattice is the winner. Thus, we recover \eqref{w1:final}. In Sec.~\ref{sec:one-sided}, we also computed $w^{(2)}$ and $w^{(3)}$, see \eqref{w2:final} and \eqref{w3:final}, and guessed the general formula \eqref{wj:final} for $w^{(j)}$ with arbitrary $j\geq 1$. 

Let $c_L$ be the probability that the agent at the leftmost site in the segment is the lucky winner in the final state. For small $L$, one can compute $c_L$ by hand:
\begin{equation}
c_1=1,\quad c_2=0, \quad c_3 = \frac{1}{4}, \quad c_4 = \frac{5}{24}
\end{equation}
etc. Generally for $L\geq 2$, the probability $u_L$ satisfies the recurrence
\begin{equation}
\label{cL:rec}
c_L =  \frac{1}{2(L-1)}\sum_{k=2}^{L-1} c_k + \frac{1}{2(L-1)}\sum_{k=1}^{L-2} c_k
\end{equation}
which is derived similarly to the recurrence \eqref{wL:rec}. Massaging the recurrence \eqref{cL:rec} we recast it into
\begin{equation}
\label{cL:rec-2}
Lc_L + \frac{1}{2}\left(c_1+c_{L-1}\right)  = \sum_{k=1}^{L} c_k 
\end{equation}
which is valid for $L\geq 1$ if we set $c_0+c_1=0$. 

Using the generating function 
\begin{equation}
\label{Cx:def}
\mathcal{C}(x) = \sum_{L\geq 1} c_{L}\, x^L
\end{equation}
together with $c_0=-1$ and $c_1=1$, we convert the recurrence \eqref{cL:rec-2} into a differential equation
\begin{equation}
\frac{d \mathcal{C}(x)}{dx} = \left(\frac{1}{x}+\frac{1}{1-x}-\frac{1}{2}\right)\mathcal{C}(x) - \frac{1}{2}\,\frac{x}{1-x}
\end{equation}
which is solved to yield the generating function 
\begin{equation}
\label{Ux:sol}
\mathcal{C}(x) = \frac{x}{1-x}\left(2e^{-x/2}-1\right)
\end{equation}
Comparing the generating functions \eqref{Wx:sol} and \eqref{Ux:sol} we conclude that
\begin{equation}
\label{bc}
c_L = 2b_L-1
\end{equation}
Using recurrences \eqref{wL:rec-2} and \eqref{cL:rec-2}, one can alternatively verify the relation \eqref{bc}. 

Since $b_\infty=e^{-1/2}$, we have $c_\infty=2e^{-\frac{1}{2}} - 1$. Equivalently, this is the probability $w_0^{(1)}$ that the first agent in the semi-infinite lattice is the lucky winner, i.e., it has not won a game. Thus
\begin{equation}
\label{u:inf}
w_0^{(1)} = 2e^{-\frac{1}{2}} - 1
\end{equation}
The agent in the left-most site played a game and won it with probability
\begin{equation}
\label{v:inf}
w_1^{(1)}  = w^{(1)}  - w_0^{(1)}  = 1-e^{-\frac{1}{2}}
\end{equation}

The virtue of the `one-sided' probabilities $w_0^{(1)}$ and $w_1^{(1)}$ characterizing the first agent in the semi-infinite lattice $\mathbb{Z}_+$ is that they allow us to determine the fractions $w_0,w_1,w_2$ describing the fate of an agent in the infinite one-dimensional lattice $\mathbb{Z}$. The WTA game proceeds independently on both sides (left and right) of the agent, and therefore the two-sided probabilities $w_0,w_1,w_2$ can be expressed through the one-sided probabilities:
\begin{equation}
\label{012}
w_0= w_0^{(1)} w_0^{(1)}, \quad w_1 = 2w_0^{(1)} w_1^{(1)} , \quad w_2= w_1^{(1)}w_1^{(1)} 
\end{equation}
Using \eqref{u:inf}--\eqref{v:inf} together with \eqref{012} we arrive at the announced results \eqref{winners:012}. 

\subsection{Maximal and minimal possible numbers of losers}
\label{subsec:min-max}

The maximal and minimal possible numbers of losers are given by \eqref{min-max}. The probability $P(N_\text{max},L)$ that the number of losers is maximal can be computed recursively.  The first loser must be an agent on the boundary of the segment with $L$ sites; this happens with probability $1/(L-1)$. The second loser must also be on the boundary of the segment with $L-1$ sites; this happens with probability $1/(L-2)$. Therefore 
\begin{equation}
\label{P:max}
P(L-1,L) = \frac{1}{(L-1)!} 
\end{equation}

The probability $P(N_\text{min},L)$ that the number of losers is minimal, $N_\text{min} = \left\lfloor \frac{L}{2}\right\rfloor$, is harder to compute recursively since the result depends on the parity of $L$. The exact answers
\begin{subequations}
\begin{align}
\label{P:min-odd}
&P(m+1,2m+1) = 2^{-m}\\
\label{P:min-even}
&P(m,2m)=\frac{2m+1}{3\cdot 2^{m-1}}
\end{align}
\end{subequations}
are derived in Sec.~\ref{subsec:GF}.

\section{Full Counting Statistics}
\label{sec:FCS}

In Sec.~\ref{subsec:av}, we determined the average number of losers in a segment. A similar calculation allows one to compute the cumulant generating function (Sec.~\ref{subsec:high}). We also determine the probabilities of reaching the final state with the minimal or maximal number of winners, and compute the generating function that encodes the exact probability distribution $P(N,L)$. We employ the same techniques as in the computation of large deviations in the RSA, see \cite{Krapivsky20}.

\subsection{Higher cumulants}
\label{subsec:high}

Consider the quantity 
\begin{equation}
\label{F:def}
F(\lambda, L) \equiv \langle e^{\lambda N}\rangle = \sum_N e^{\lambda N} P(N,L)
\end{equation}
where $P(N,L)$ is the probability to have $N$ losers. The standard relation
\begin{equation}
\ln \langle e^{\lambda N}\rangle = \sum_{n\geq 1} \frac{\lambda^n}{n!}\, \langle\!\langle N^n\rangle\!\rangle
\end{equation}
then gives all the cumulants: the average $\langle\!\langle N \rangle\!\rangle = \langle N\rangle$, the variance $\langle\!\langle N^2\rangle\!\rangle = \langle N^2\rangle - \langle N\rangle^2$, etc. 

The function $F(\lambda, L) \equiv \langle e^{\lambda N}\rangle$ grows exponentially:
\begin{equation}
\label{U:def}
\ln F(\lambda, L) \simeq LU(\lambda) \quad\text{when}\quad L\to\infty
\end{equation}
Expanding the cumulant generating function $U(\lambda)$ into the Taylor series we can read off the cumulants: 
\begin{equation}
\label{U:cum}
U(\lambda)=\sum_{n\geq 1} \frac{\lambda^n}{n!}\, U_n, \qquad\langle\!\langle N^n\rangle\!\rangle = L U_n
\end{equation}

To determine $F(\lambda, L)$ we proceed in the same way as in the derivation of the average and obtain a recurrence
\begin{eqnarray}
\label{FL:rec}
F(\lambda, L) &=& \frac{e^\lambda}{L-1}\sum_{k=2}^{L-1} F(\lambda, k-1) F(\lambda, L-k)\nonumber\\
&+& \frac{e^\lambda}{L-1}\,F(\lambda,L-1)
\end{eqnarray}
The terms in the sum on the right-hand side correspond to situations with the first loser in bulk sites $k=2,\ldots,L-1$; the losers on the intervals to the left and  right of site $k$ arise independently. (The one-dimensional nature of the problem is crucial for this property.) The last term on the right-hand side of \eqref{FL:rec} describes situations in which the first loser is at a boundary site. 

The recurrence \eqref{FL:rec} applies to all $L\geq 2$ if we use the known value 
\begin{equation}
\label{FL:IC}
F(\lambda, 1)=1
\end{equation}
at $L=1$ as an initial condition. As a check, \eqref{FL:rec}--\eqref{FL:IC} reproduce the known values of $ F(\lambda, L)$ for small $L$:
\begin{equation*}
F(\lambda, 2)=e^\lambda, \quad  F(\lambda, 3)=\tfrac{1}{2}e^\lambda + \tfrac{1}{2}e^{2\lambda}
\end{equation*}

To solve the recurrence \eqref{FL:rec} we introduce the generating function
\begin{equation}
\mathcal{F}(\lambda, x) = \sum_{L\geq 1}  F(\lambda, L)\, x^{L-1}
\end{equation}
Multiplying \eqref{FL:rec} by $(L-1)x^{L-2}$ and summing over all $L\geq 2$ we obtain
\begin{equation}
\label{Phi:eq}
\frac{d \mathcal{F}(\lambda, x)}{d x} = e^\lambda \mathcal{F}(\lambda, x)\left[1+x\mathcal{F}(\lambda, x)\right]
\end{equation}
To solve this Bernoulli equation, we use the standard substitution $\mathcal{F} = 1/u$. Equation \eqref{Phi:eq} turns into a linear differential equation $\frac{du}{dx}=-e^\lambda x - e^\lambda u$. Solving this equation and using the initial condition $\mathcal{F}(\lambda, 0)=1$ we obtain
\begin{equation}
\label{Phi:sol}
\mathcal{F}(\lambda, x) = \frac{1}{e^{-\lambda}-x+(1-e^{-\lambda})\exp[-e^\lambda x]}
\end{equation}
The generating function $\mathcal{F}(\lambda, x)$ has a simple pole located at $x=y=y(\lambda)$ which is found from
\begin{equation}
\label{y:def}
e^{-\lambda}-y+(1-e^{-\lambda})\exp[-e^\lambda y] = 0
\end{equation}
The corresponding Laurent series reads 
\begin{equation}
\mathcal{F}(\lambda, x) = \frac{C}{y-x}  + \frac{C-1}{2 y}+ O(x-y)
\end{equation}
with $C=y^{-1}e^{-\lambda}$. To match this simple pole  the function $F(\lambda, L)$ must exhibit the following asymptotic ($L\gg 1$) behavior 
\begin{equation}
\label{FU}
F(\lambda, L) \simeq e^{-\lambda}\, e^{(L+1)U(\lambda)}, \quad U(\lambda) = -\ln y(\lambda)
\end{equation}
This asymptotic implies \eqref{U:def} and gives the cumulant generating function $U(\lambda)$. Using \eqref{y:def} and \eqref{FU} we express $U(\lambda)$ as a solution of
\begin{equation}
\label{U:sol}
e^{-\lambda}-e^{-U}+\left(1-e^{-\lambda}\right)\exp\!\left[-e^{\lambda-U}\right] = 0
\end{equation}

\begin{figure}[ht]
\begin{center}
\includegraphics[width=0.48\textwidth]{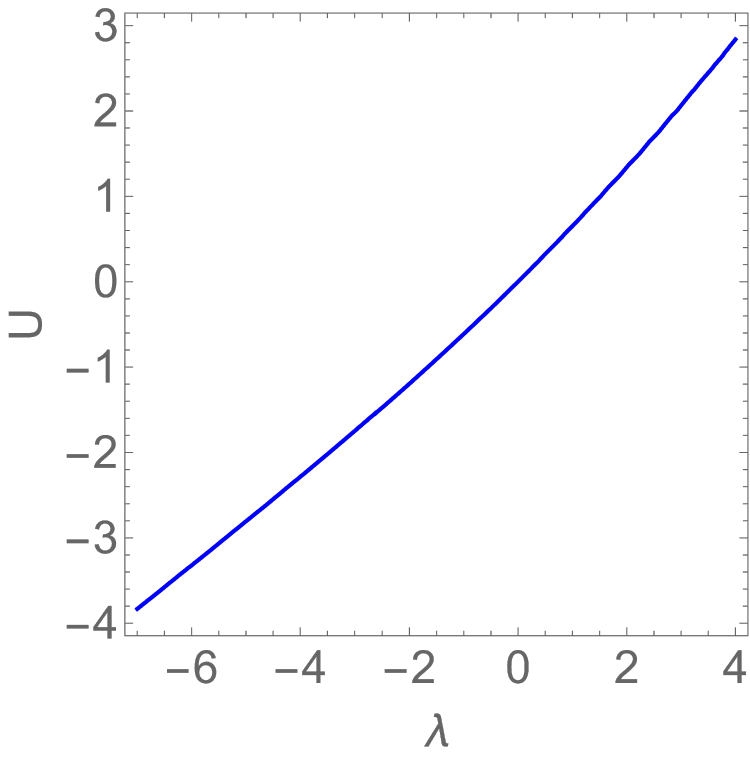}
\caption{The plot of the cumulant generating function $U(\lambda)$. The expansion of $U(\lambda)$ at $\lambda=0$ yields the cumulants.}
\label{fig:U-winner}
  \end{center}
\end{figure}

The cumulant generating function $U(\lambda)$ implicitly determined by Eq.~\eqref{U:sol} is plotted in Fig.~\ref{fig:U-winner}. Using \eqref{U:cum} and  \eqref{U:sol}, we confirm already known leading behavior of the average, $\lim_{L\to\infty}L^{-1}\langle N\rangle = 1-e^{-1}$, and determine leading behavior of the variance
\begin{equation}
\label{var:lim}
\lim_{L\to\infty} L^{-1} \langle\!\langle N^2\rangle\!\rangle = 3e^{-2}-e^{-1}
\end{equation}
Higher cumulants also scale linearly with $L$. Using \eqref{U:cum} and  \eqref{U:sol}, one can extract the leading behavior of any cumulant. For instance, 
\begin{equation*}
\begin{split}
\frac{\langle\!\langle N^3\rangle\!\rangle}{L} & = -17e^{-3}+9e^{-2}-e^{-1}\\
\frac{\langle\!\langle N^4\rangle\!\rangle}{L} & = 142e^{-4} -102 e^{-3}+21 e^{-2}-e^{-1}\\
\frac{\langle\!\langle N^5\rangle\!\rangle}{L} & = -1569 e^{-5}+1420 e^{-4} -425 e^{-3}+45 e^{-2}-e^{-1} \\
\frac{\langle\!\langle N^6\rangle\!\rangle}{L} & = 21576 e^{-6} - 23535 e^{-5}+9230 e^{-4} -1530 e^{-3}\\
&+93 e^{-2}-e^{-1}
\end{split}
\end{equation*}

The ratios $\langle\!\langle N^n\rangle\!\rangle/\langle N\rangle$ are the Fano factors \cite{Fano}:
\begin{equation*}
\label{Fano}
\begin{split}
\frac{\langle\!\langle N^2\rangle\!\rangle}{\langle N\rangle} & = \frac{3e^{-1}-1}{e-1} = 0.060\,315\,090\,224\ldots \\
\frac{\langle\!\langle N^3\rangle\!\rangle}{\langle N\rangle} & = \frac{9e^{-1}-17e^{-2}-1}{e-1}=0.005\,944\,982\,570\ldots \\
\frac{\langle\!\langle N^4\rangle\!\rangle}{\langle N\rangle} & =\frac{142e^{-3} -102 e^{-2}+21 e^{-1}-1}{e-1}=-0.0052\ldots
\end{split}
\end{equation*}
These values demonstrate that the fluctuations differ substantially from Poisson statistics for which all Fano factors are equal to one.

\subsection{Exact probability distribution $P(N,L)$}
\label{subsec:GF}

Using the same arguments as in deriving \eqref{AL:rec} and \eqref{FL:rec} one finds that the probability distribution $P(N,L)$ satisfies a recurrence relation
\begin{eqnarray}
\label{PNL:rec}
(L-1) P(N,L) &=& \sum_{k=2}^{L-1}\sum_{i+j=N-1}P(i,k-1)P(j,L-k) \nonumber\\
 &+& P(N-1,L-1) 
\end{eqnarray}
for $L\geq 2$. We know  
\begin{equation}
\label{PN12}
P(N,1)=\delta_{N,0}, \qquad P(N,2)=\delta_{N,1}
\end{equation}
and using \eqref{L=3}--\eqref{L=5} one finds $P(N,L)$ for $L=3,4,5$:
\begin{equation}
\label{345}
\begin{split}
P(N,3) &=\tfrac{1}{2}\delta_{N,1}+\tfrac{1}{2}\delta_{N,2}\\
P(N,4) &=\tfrac{5}{6}\delta_{N,2}+\tfrac{1}{6}\delta_{N,3}\\
P(N,5) &=\tfrac{1}{4}\delta_{N,2}+\tfrac{17}{24}\delta_{N,3}+\tfrac{1}{24}\delta_{N,4}
\end{split}
\end{equation}
These results are  consistent with \eqref{PNL:rec} thereby providing a useful check of the recurrence \eqref{PNL:rec}. We also notice that Eq.~\eqref{P:max} is confirmed in these examples. 

Introducing the generating function
\begin{equation}
\label{Pxy}
\mathcal{P}(x,y)=\sum_{N\geq 0}\sum_{L\geq 0}P(N,L) x^N y^{L-1}
\end{equation}
we recast the recurrence \eqref{PNL:rec} into 
\begin{equation}
\label{P:ODE}
\frac{\partial \mathcal{P}}{\partial y} = x\mathcal{P}[1+y\mathcal{P}]
\end{equation}
Solving this differential equation subject to the boundary condition $\mathcal{P}(x,0)=1$ which follows from \eqref{PN12} one gets 
\begin{equation}
\label{Pxy:sol}
\mathcal{P}(x,y)=\frac{x}{1-xy+(x-1)e^{-xy}}
\end{equation}
Expanding the generating function \eqref{Pxy:sol} one confirms \eqref{345} and computes $P(N,L)$ for $L\geq 6$:
\begin{equation*}
\label{6-12}
\begin{split}
P(N,6) &=\tfrac{7}{12}\delta_{N,3}+\tfrac{49}{120}\delta_{N,4}+\tfrac{1}{5!}\delta_{N,5}\\
P(N,7) &=\tfrac{1}{8}\delta_{N,3}+\tfrac{25}{36}\delta_{N,4}+\tfrac{43}{240}\delta_{N,5}+\tfrac{1}{6!}\delta_{N,6}\\
P(N,8) &=\tfrac{3}{8}\delta_{N,4}+\tfrac{101}{180}\delta_{N,5}+\tfrac{107}{1680}\delta_{N,6}+\tfrac{1}{7!}\delta_{N,7}\\
P(N,9) &=\tfrac{1}{16}\delta_{N,4}+\tfrac{55}{96}\delta_{N,5}+\tfrac{199}{576}\delta_{N,6}+\tfrac{769}{8!}\delta_{N,7}\\
           & +\tfrac{1}{8!}\delta_{N,8}\\
P(N,10) &=\tfrac{11}{48}\delta_{N,5}+\tfrac{2563}{4320}\delta_{N,6}+\tfrac{10439}{60480}\delta_{N,7}+\tfrac{1793}{9!}\delta_{N,8} \\
             &+\tfrac{1}{9!}\delta_{N,9}\\
P(N,11) &=\tfrac{1}{32}\delta_{N,5}+\tfrac{41}{96}\delta_{N,6}+\tfrac{8083}{17280}\delta_{N,7} +\tfrac{11003}{151200}\delta_{N,8}\\
             &+\tfrac{4097}{(10)!}\delta_{N,9}+\tfrac{1}{(10)!}\delta_{N,10}\\
P(N,12) &=\tfrac{13}{96}\delta_{N,6}+\tfrac{2327}{4320}\delta_{N,7} +\tfrac{36179}{120960}\delta_{N,8} +\tfrac{377}{14175}\delta_{N,9}\\
             &+\tfrac{9217}{(11)!}\delta_{N,10}+\tfrac{1}{(11)!}\delta_{N,11} \\
P(N,13) &=\tfrac{1}{64}\delta_{N,6}+\tfrac{343}{1152}\delta_{N,7} +\tfrac{13391}{25920}\delta_{N,8} +\tfrac{41827}{259200}\delta_{N,9}\\
             &+\tfrac{26723}{3110400}\delta_{N,10}+\tfrac{20481}{(12)!}\delta_{N,11}+\tfrac{1}{(12)!}\delta_{N,12}
\end{split}
\end{equation*}
etc. These results suggest that the probabilities of attaining a jammed configuration with the minimal number of losers are given by \eqref{P:min-odd}--\eqref{P:min-even}. 

Let us derive a simpler result, Eq.~\eqref{P:min-odd}, for the probability of the minimal number of losers in segments with an odd number of sites. We begin with an example. For the segment with 15 sites, the minimal number of losers is 7, and the final state is
\begin{equation}
\label{min-7}
\blacktriangleleft\,\circ\,\bullet\,\circ\,\bullet\,\circ\,\bullet\,\circ\,\bullet\,\circ\,\bullet\,\circ\,\,\,\blacktriangleright
\end{equation}
where we again use special notations, $\blacktriangleleft$ and $\blacktriangleright$, for the agents on the boundary, that happened to be winners in \eqref{min-7}. A similar outcome for the segment $[1,\ldots,2m+1]$ emerges when, in every game, the agent with an even label loses. The number of games in this situation is $m$, so the probability is $2^{-m}$ as stated in \eqref{P:min-odd}. 

We do not give the corresponding derivation of \eqref{P:min-even} for segments with an even number of sites. The complication is that there are many outcomes with a minimal number of losers. For instance, there are four outcomes
\begin{equation*}
\begin{split}
&\blacktriangleleft\,\circ\,\bullet\,\circ\,\,\bullet\,\,\vartriangleright  \\
&\blacktriangleleft\,\circ\,\bullet\,\circ\,\,\circ\,\,\blacktriangleright \\
&\blacktriangleleft\,\circ\,\circ\,\bullet\,\,\circ\,\,\blacktriangleright \\
&\vartriangleleft\,\bullet\,\circ\,\bullet\,\,\circ\,\,\blacktriangleright 
\end{split}
\end{equation*}
for $L=6$, and generally $m+1$ outcomes when $L=2m$. 

It is easier to derive the probability of attaining jammed configurations with the maximal number of losers than with the minimal number of losers. The probability of attaining a jammed configuration with penultimate number of losers
\begin{equation}
\label{PL2}
P(L-2,L) = \frac{1+(L-3)2^{L-2}}{(L-1)!} 
\end{equation}
is also easy to deduce from the recurrence \eqref{PNL:rec}:

\section{Infinite $b$-ary trees}
\label{sec:Bethe}

\begin{figure}[ht]
\begin{center}
\includegraphics[width=0.48\textwidth]{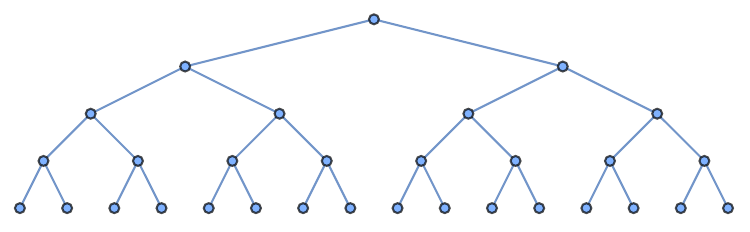}
\caption{A binary tree. Only the first four generations are shown: The root constitutes generation zero, two children of the root constitute generation one, four grandchildren constitute generation two, etc. }
\label{fig:binary}
  \end{center}
\end{figure}

An infinite binary tree has a root vertex linked to two `children' vertices; see Fig.~\ref{fig:binary}. Each vertex different from the root has three neighbors: a parent vertex and two child vertices. More generally, in an infinite $b$-ary tree, each vertex has $b$ children: The root has degree $b$ while all other vertices have degree $b+1$. In the $d$-regular graph, all vertices have the same degree $d$. An infinite regular tree is also often called a Bethe lattice, and the degree is called the coordination number. 

For the WTA process on the infinite $b$-ary tree, one natural quantity is the probability $\omega(t)$ that the agent at the root has not lost a game during the time interval $(0,t)$. For the agent far away from the root, the probability $w(t)$ coincides with the fraction of winners (at time $t$) on the Bethe lattice with coordination number $d=b+1$. 

We begin with the root. Keeping the root and one of its neighbors with an emanating branch, we denote by $\Omega(t)$ the probability that the root in this reduced tree is the winner. Note that $\omega(t)$ on the original tree and $\Omega(t)$ on the reduced tree are related by $\omega = \Omega^b$, reflecting the independence of branches. Another relation 
\begin{equation}
\label{Omega:eq}
\frac{d\Omega}{dt} = - \frac{1}{2}\,e^{-t}\omega
\end{equation}
contains the probability density $e^{-t}$ that the first game between the agent in the root and the child is played during the time interval $(t,t+dt)$, the probability $\omega$ that the child has not yet lost a game, and the probability $\frac{1}{2}$ that the agent in the root loses a game. 

Using the modified time variable and $\omega=\Omega^b$ we recast \eqref{Omega:eq} into $\frac{d\Omega}{d\tau} = - \frac{1}{2}\,\Omega^b$ which is integrated to yield
\begin{equation}
\label{Omega:sol}
\Omega(\tau)=\left[1+\frac{b-1}{2}\,\tau\right]^{-\frac{1}{b-1}}
\end{equation}
from which 
\begin{equation}
\label{omega:sol}
\omega(\tau)=\left[1+\frac{b-1}{2}\,\tau\right]^{-\frac{b}{b-1}}
\end{equation}
In the final state ($\tau=1$)
\begin{equation}
\label{omega:final}
\Omega=\left(\frac{2}{b+1}\right)^\frac{1}{b-1}\,, \qquad 
\omega=\left(\frac{2}{b+1}\right)^\frac{b}{b-1}
\end{equation}

The agent in a vertex far from the root is a winner with probability $w=\Omega^{b+1}$. In the final state
\begin{equation}
\label{w:final}
w = \left(\frac{2}{b+1}\right)^\frac{b+1}{b-1}
\end{equation}
The probabilities $\omega$ and $w$ are decreasing functions of $b$.

\begin{figure}[ht]
\begin{center}
\includegraphics[width=0.48\textwidth]{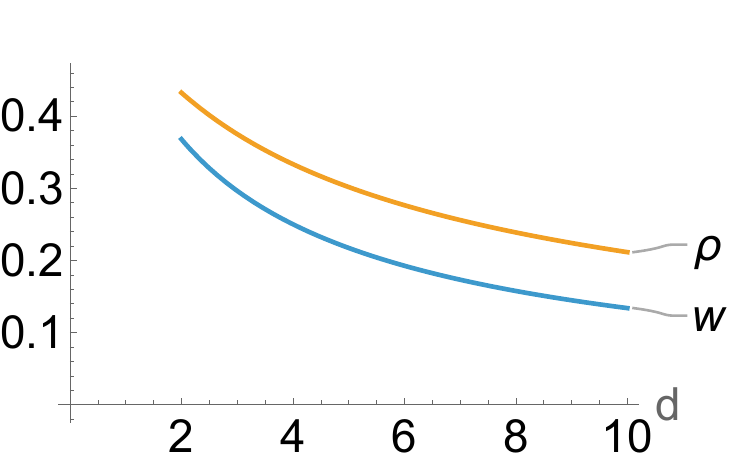}
\caption{Winner density $w$ for the WTA process and the jamming density $\rho$ for the simplest RSA process on the Bethe lattice, as functions of the coordination number $d$. The exact expressions are given by Eqs.~\eqref{w:Bethe}--\eqref{rho:Bethe}. These densities are decreasing functions of $d$ with asymptotics $w\simeq \frac{2}{d}$ and $\rho\simeq d^{-1}\log d$ when $d\gg 1$.}
\label{fig:rho-w}
  \end{center}
\end{figure}

The $b$-ary tree far away from the root is indistinguishable from the infinite Bethe lattice with coordination number $d=b+1$. Thus, the fraction of winners $w(d)$ on the Bethe lattice with coordination number $d$ is 
\begin{equation}
\label{w:Bethe}
w(d) = \left(\frac{2}{d}\right)^\frac{d}{d-2}
\end{equation}
For comparison, the jamming density $\rho(d)$ for the simplest RSA process on the Bethe lattice is \cite{Pippenger,Percus91}
\begin{equation}
\label{rho:Bethe}
\rho(d) =\frac{1}{2}\left[1-(d-1)^{-\frac{2}{d-2}}\right]
\end{equation}
Using exact expressions \eqref{w:Bethe}--\eqref{rho:Bethe} one can verify that $\rho(d)>w(d)$ for all $d\geq 2$; see also Fig.~\ref{fig:rho-w}.

\section{Discussion}
\label{sec:disc}

We have obtained exact results for the WTA process on infinite and semi-infinite one-dimensional lattices, finite segments, and Bethe lattices. On the one-dimensional lattice, the final fraction of winners is $e^{-1}$, and we have determined the distribution of loser domains and the equal-time pair correlation function. For finite segments, we obtained the full counting statistics of the number of losers, including all cumulants and the probabilities of attaining the minimum and maximum possible numbers. We also determined boundary effects and extended the analysis to Bethe lattices.

Many quantities remain to be determined and may admit an analytical treatment. For example, the pair correlation function computed in Sec.~\ref{sec:corr} ignores the states of the winners. A more detailed pair correlation function is $C_j^{(a,b)}$, with $a,b\in\{0,1,2\}$  referring to the number of wins by the agent on the left and right. For instance, 
\begin{equation}
C_2^{(0,1)} = \left\langle O_0^{(0)} O_2^{(1)}\right\rangle - w_0 w_1
\end{equation}
The correlators $\left\langle O_0^{(a)} O_2^{(b)}\right\rangle$ vanish when $a=b=0$ or $a=b=2$: The lucky winners, as well as the winners in two games, cannot be separated by just one loser. Thus in the first nontrivial case of $j=2$, 
\begin{equation}
C_2^{(0,0)}= - w_0^2, \qquad C_2^{(2,2)} = - w_2^2
\end{equation}
The remaining $C_2^{(a,b)}$ are unknown. One can also try to compute a more detailed domain size distribution $\Pi_s^{(a,b)}$. 

The simplicity of the domain size distribution, Eq.~\eqref{Pi_s}, hints that the distribution 
\begin{equation}
\label{Pi-ss:def}
\Pi_{s_1,s_2} = \text{Prob}\big[\bullet \underbrace{\circ \cdots \circ}_{s_1} \bullet  \underbrace{\circ \cdots \circ}_{s_2} \bullet\big]
\end{equation}
describing the correlation between adjacent domains of losers can be simple. To compute $\Pi_{s_1,s_2}$, one can employ the same approach as in Sec.~\ref{subsec:domain} and study the evolution of the densities $P_{s_1,s_2}(m_1,m_2,m_3;t)$ of patterns
\begin{equation*}
\underbrace{\bullet \cdots \bullet}_{m_1} \underbrace{\circ \cdots \circ}_{s_1}\underbrace{\bullet \cdots \bullet}_{m_2}\underbrace{\circ \cdots \circ}_{s_2} \underbrace{\bullet \cdots \bullet}_{m_3} 
\end{equation*}

More generally, the densities $P_{\bf s}({\bf m};t)$ with $k$ clusters ${\bf s}=(s_1,\ldots,s_k)$ of losers, $s_i\geq 1$, and $(k+1)$ clusters of active agents ${\bf m}=(m_1,\ldots,m_{k+1})$, with $m_j\geq 1$, can perhaps be tractable by extending the technique of Sec.~\ref{subsec:domain}. In the final state, only the densities $\Pi_{\bf s}\equiv P_{\bf s}({\bf 1};\infty)$ with ${\bf m}={\bf 1}=(1,\ldots,1)$ are non-vanishing.

The WTA process is tractable on the Bethe lattices and $b$-ary trees.  The analytical treatment of the WTA process on graphs with numerous short cycles is out of reach: An exact analytical treatment appears impossible for the triangular, square, and hexagonal lattices in two dimensions. Quasi-one-dimensional graphs constitute a rare exception. For instance, similarly the simplest RSA process \cite{ladder92,ladder15,ladder24}, solving the WTA process on the ladders should be feasible. It would also be useful to analyze the WTA process on Erd\H{o}s-R{\'e}nyi random graphs. 

We finally mention conjectural formulae for the fractions of winners on the hypercubes $\mathbb{H}^d$ and the hypercubic lattices $\mathbb{Z}^d$ in the limit of large spatial dimension:
\begin{equation}
\label{w:hyper}
w_{\mathbb{H}^d} \simeq \frac{2}{d}\,, \qquad w_{\mathbb{Z}^d} \simeq \frac{1}{d}
\end{equation}
To appreciate the first asymptotic, we note that the hypercube $\mathbb{H}^d$, when treated as a graph, has $2^d$ vertices, resembling the Bethe lattice with coordination number $d$. The hypercube has cycles, but they are rare when $d\gg 1$. Hence, one anticipates that the fraction of winners coincides with the leading asymptotic of the exact answer \eqref{w:Bethe} for the Bethe lattice. Similarly, $\mathbb{Z}^d$ resembles the Bethe lattice with coordination number $2d$ and the cycles also become rare as $d\to\infty$. The conjectural behaviors \eqref{w:hyper} follow from \eqref{w:Bethe}. 

To gain deeper insight and stronger justification for \eqref{w:hyper}, it would be valuable to develop an approach that provides more solid support of Eq.~\eqref{w:hyper}. Such an approach could lead to the asymptotic expansions
\begin{equation*}
\begin{split}
w_{\mathbb{H}^d} & = A_1 d^{-1}+A_2 d^{-2} + \ldots \\
w_{\mathbb{Z}^d} & = B_1 d^{-1}+B_2 d^{-2} + \ldots
\end{split}
\end{equation*}
allow to fix the leading terms, $A_1=2$ and $B_1=1$, and potentially compute the amplitudes $A_2$ and $B_2$ for the subleading terms. 

For the simplest RSA, the asymptotic behaviors of the jamming density, $\rho_{\mathbb{H}^d} \simeq d^{-1}\log d$ and $\rho_{\mathbb{Z}^d} \simeq (2d)^{-1}\log d$. likewise follow from the exact answer \eqref{rho:Bethe}.  These behaviors appear to be exact leading asymptotics \cite{Baram89,Baram21}.

\bigskip
\appendix
\section{Infinite linear graphs with many arms}
\label{ap:arms}

Here, we briefly consider a family of infinite linear graphs parametrized by the number $r$ of arms. The graph with $r=1$ is the semi-infinite lattice $\mathbb{Z}_+$ treated as a rooted graph, with the root being the boundary vertex. Taking $r$ copies of $\mathbb{Z}_+$ and identifying the roots we obtain a linear graph $\mathbb{Z}^{(r)}$ with $r$ arms. Thus, $\mathbb{Z}^{(1)}=\mathbb{Z}_+$ and $\mathbb{Z}^{(2)}=\mathbb{Z}$, while $\mathbb{Z}^{(r)}$ with $r\geq 3$ arms are new infinite linear graphs. Generally, the union of two rooted graphs $G_1$ and $G_2$ obtained by identifying their roots, $G_1\vee G_2$, is known as the wedge sum. Thus
\begin{equation}
\mathbb{Z}^{(r)} = \underbrace{\mathbb{Z}_+\vee \cdots\vee \mathbb{Z}_+}_r
\end{equation}

For the WTA process on $\mathbb{Z}^{(r)}$, let $\omega^{(r)}(t)$ be the probability that the agent in the root is active at time $t$. (We write $\omega^{(r)}$ instead of $w^{(r)}$; the latter notation was used for the probability that the agent with label $r$ on $\mathbb{Z}_+$ is the winner.) We already know that $\omega^{(1)}(t)=e^{-\tau/2}$, since $\omega^{(1)}=w^{(1)}$, and the latter is known \eqref{w1:sol}. We also know $\omega^{(2)}(t)=e^{-\tau}$, denoted earlier as $w(t)$ as it corresponds to an arbitrary agent in $\mathbb{Z}$. We see that $\omega^{(2)}=\big[\omega^{(1)}\big]^2$, and generally
\begin{equation}
\omega^{(r)}(t)=\big[\omega^{(1)}(t)\big]^r = e^{-r\tau/2}
\end{equation}
Therefore, the agent in the root is the winner with probability
\begin{equation}
\omega^{(r)}  = \omega^{(r)}(t=\infty) = e^{-r/2}
\end{equation}

We already computed the probability $2e^{-1/2} - 1$ [respectively $1-e^{-1/2}$] that the agent in the root of $\mathbb{Z}_+$ wins zero (respectively one) game, see \eqref{u:inf}--\eqref{v:inf}. The agent at the root of $\mathbb{Z}^{(r)}$ wins $j$ games with probability
\begin{equation}
\omega_j^{(r)} = \binom{r}{j}\left(2e^{-\frac{1}{2}} - 1\right)^{r-j}\left(1-e^{-\frac{1}{2}}\right)^j
\end{equation}

\bibliography{references-packing}

\end{document}